\documentclass[conference]{IEEEtran}
\usepackage{cite}
\usepackage{amsmath,amssymb,amsfonts}
\usepackage{graphicx}
\usepackage{textcomp}
\usepackage{xcolor}

\usepackage{booktabs}
\usepackage{tabularx}
\usepackage{algorithm}
\usepackage{algpseudocode}

\usepackage{array}
\usepackage{multirow}
\usepackage{subcaption}

\newcommand{\code}[1]{\texttt{\small #1}}

\def\BibTeX{{\rm B\kern-.05em{\sc i\kern-.025em b}\kern-.08em
    T\kern-.1667em\lower.7ex\hbox{E}\kern-.125emX}}
\begin{document}

\title{Runtime Authorization Consistency Checking for MCP-based Agentic Workflows}

\author{
\IEEEauthorblockN{Aiyao Zhang\textsuperscript{1,2},
Xiaodong Lee\textsuperscript{1,3,4,\S},
Zhixian Zhuang\textsuperscript{1,2},
Botao Peng\textsuperscript{1,3}}
\IEEEauthorblockA{\textsuperscript{1}Institute of Computing Technology, Chinese Academy of Sciences\\
\textsuperscript{2}University of Chinese Academy of Sciences\\
\textsuperscript{3}Fuxi Institution\\
\textsuperscript{4}Center for Internet Governance, Tsinghua University\\
\textsuperscript{\S}Corresponding author\\
\{zhangaiyao22z, xl, zhuangzhixian22s, pengbotao\}@ict.ac.cn}
}

\maketitle

\begin{abstract}
Agentic systems increasingly fulfill user requests through multi-step tool workflows over files, services, and external resources. In these workflows, isolated per-call checks can miss a workflow-level failure: each call may be locally admissible, but the sequence can exceed the authorization boundary established for the session. We identify this failure mode ``authorization drift". To address this problem, we present Runtime Authorization Consistency Checking (RAC), a lightweight guard at the controller-side tool-call boundary. RAC treats authorization as runtime state carried by accepted workflow steps. For each pending action, it reconstructs a trusted authorization event from controller-observed metadata and admits the call only when it remains no more permissive than the basis inherited through accepted lineage. Rejected steps are excluded from lineage, so later continuations can draw support only from accepted workflow history. Our evaluation shows that RAC reduces missed authorization drift across both controlled and planner-generated workflows. On the 1,248-workflow TraceBench suite, RAC has no missed-block cases, while the strongest Static+History baseline misses 509 of 1,008 oracle-BLOCK workflows. On a high-confidence observable subset of blind LLM-generated plans, RAC reaches 92.8\% block recall, compared with 68.8\% for the strongest baseline. In the real MCP filesystem planner replay, RAC stops nine unsafe continuations before server execution, with sub-millisecond p99 checking latency.
\end{abstract}

\section{Introduction}
\subsection{Background and Problem}\label{sec:background}

Agent-based systems increasingly use large language models as planners that decompose user requests, invoke external tools, and adapt subsequent actions based on intermediate observations~\cite{yao2022react,schick2023toolformer}. With communication protocols such as the Model Context Protocol (MCP)~\cite{mcp}, these agents can access external tools through structured interfaces, turning a user-level request into a runtime-expanded workflow~\cite{mcp_review,he2025security}. In more complex deployments, parts of this workflow may further be delegated to other agents or specialized components~\cite{south2025identity}.

Existing defenses have made important progress on securing LLM-integrated agents. Prior work has studied prompt-injection defenses~\cite{greshake2023not}, unsafe tool-use evaluation~\cite{ruan2024identifying}, task-alignment checks~\cite{jia2025task}, and controlled benchmarks for tool-using agents~\cite{debenedetti2024agentdojo}. These defenses provide important foundations for protecting model-facing inputs, constraining unsafe tool behavior, and evaluating whether agents remain aligned with the intended task.

However, a less explored \emph{authorization issue} emerges when a user-level task is executed as a sequence of tool calls. In multi-step MCP-based workflows, each call may be locally permitted, while the session-level authority remains tied to the original task scope~\cite{mcp_review}. Recent multi-agency cybersecurity guidance~\cite{agentic_ai_guidance_2026} warns against broad or unrestricted agentic access and identifies privilege compromise and scope creep as key deployment risks. In such workflows, scope creep may arise gradually, as individually permissible calls accumulate into a workflow that goes beyond the authority granted for the session~\cite{hardy1988confused}. This \emph{workflow-level authorization gap} motivates the problem studied in this paper.

\begin{figure}[t]
  \centering
  \includegraphics[width=\columnwidth]{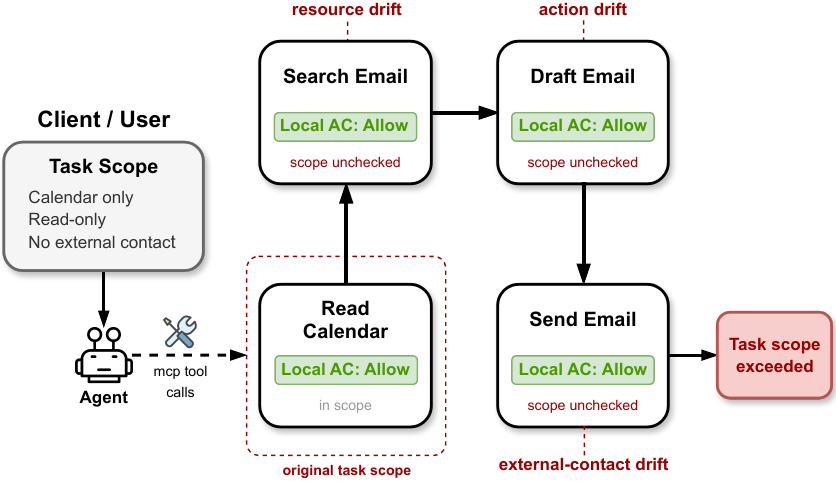}
  \caption{Authorization drift in a scheduling workflow. Locally allowed calls compose into a sequence that exceeds the initial task scope.}
  \label{fig:task-scope-drift}
\end{figure}

\subsection{Motivation}\label{sec:motivation}

Figure~\ref{fig:task-scope-drift} illustrates a representative scheduling workflow. An enterprise assistant has standing access to calendar, email, and internal messaging tools, but the user authorizes only a narrow task: find available meeting slots by inspecting the calendar, without reading related communications, modifying data, or contacting others. The assistant first retrieves calendar availability and identifies candidate slots, which is permitted by the platform and consistent with the task scope. The planner may then continue by searching related emails for meeting context, inferring a conflict, and sending a rescheduling message through an internal communication channel.

The violation appears only when the workflow is viewed as a whole. Although the later calls may still be locally executable under existing tool permissions, they are no longer justified by the original authorization boundary: the user authorized a calendar-based availability lookup, not email inspection or outbound communication. The workflow has therefore moved beyond the task scope established at the start.

We refer to this workflow-level failure as \emph{authorization drift}. A continuation can remain executable under local tool permissions while losing support from the authority carried forward by accepted workflow history. The failure unfolds across time: later steps inherit data, context, or apparent justification from earlier outputs and gradually move beyond the authorized task. The core challenge is to preserve authorization consistency as the workflow evolves, rather than treating authorization as a one-time entry decision or a set of isolated per-call checks.

\begin{figure}[t]
    \centering
    \includegraphics[width=\columnwidth]{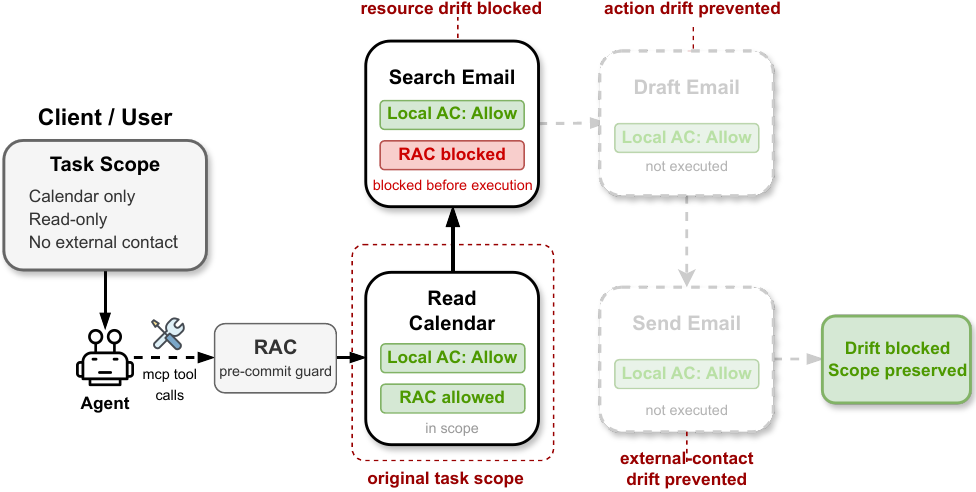}
    \caption{RAC prevents authorization drift at the pre-commit boundary. The email-search continuation is blocked after the calendar read, so downstream draft/send steps are not reached.}
    \label{fig:rac-guarded-example}
\end{figure}

\subsection{Our Solution}

We present Runtime Authorization Consistency (RAC) Checking, a lightweight pre-commit guard for MCP-based agent workflows. RAC is placed at the workflow controller before proposed tool calls are issued to external tools. It treats authorization as runtime state carried through accepted workflow history and checked at each proposed continuation. A pending call is allowed only when it remains justified by the authorization basis inherited from prior accepted steps; otherwise, it is blocked before execution and cannot become support for later continuations. RAC therefore preserves the task-level authorization boundary as the workflow evolves, while leaving MCP servers and tool implementations unchanged.

Figure~\ref{fig:rac-guarded-example} illustrates how RAC fits into the same MCP tool-call flow. In the calendar-availability example, RAC allows the initial calendar read but blocks the email-search continuation at the pre-commit boundary, so the downstream draft and send steps are not executed. Our implementation realizes RAC as a controller-side runtime guard, and our evaluation shows that this workflow-level check catches drift missed by simpler baselines, blocks unsafe continuations before MCP server execution, and adds only low checking overhead.

The key contributions of this paper are as follows:

\begin{itemize}
    \item We identify \emph{authorization drift} as an authorization failure in MCP agent workflows, where a sequence of locally admissible tool calls can move beyond the authorization boundary established for a session.

    \item We formulate runtime authorization consistency as a pre-commit safety property over accepted workflow history, requiring each pending call to remain no more permissive than the authorization basis inherited from prior accepted steps.

    \item We design and implement \emph{Runtime Authorization Consistency (RAC)} Checking, a controller-side guard that uses trusted runtime metadata, verified lineage, and inherited authorization state to block unsupported continuations before they reach external tools.

    \item We evaluate RAC with TraceBench, practical baselines, blind LLM-plan replay, a real MCP filesystem case study, and latency measurements. The results show that RAC reduces missed blocks in both controlled and planner-generated workflows, stops unsafe actions before server execution, and adds low checking overhead.
\end{itemize}

\section{Problem Formulation and Threat Model}
\label{sec:problem-threat}

\subsection{Security Objective}
\label{sec:security-objective}
We use an authorization basis to capture the runtime authority associated with an accepted workflow step. The basis summarizes who is acting, what authority remains available, and which verified evidence supports the resources carried forward by the workflow. It is initialized from the session grant and tightened only through accepted steps whose security-relevant fields are constructed from trusted runtime evidence.

A monitored MCP-based workflow is modeled as $W=\langle e_1,e_2,\ldots,e_n\rangle$, where each $e_i$ is an authorization-relevant event observed at a controller-mediated pre-commit boundary. Each workflow session starts from an authenticated subject and an initial grant envelope $G$ supplied by a trusted authorization layer. The grant defines the initial authorization boundary against which subsequent workflow steps are evaluated.

Authorization drift arises when a pending event goes beyond the basis carried forward by the accepted history, even though the event still appears locally executable. Drift is therefore defined by the relationship between a pending continuation and the authorization boundary carried forward by accepted predecessor steps. The concrete drift patterns studied in this paper are instantiated by the rule families in Section~\ref{sec:precommit-checking} and the TraceBench templates in Section~\ref{sec:evaluation}.

For a pending event $e_i$, let $pred_i$ denote the predecessor selected by RAC, where the distinguished predecessor $\bot$ is used for an initial grant-bound event. Let $B_i^{-}$ denote the authorization basis available at its pre-commit boundary. For the first authorization-relevant event, this basis is initialized from $G$; for subsequent events, it is loaded through controller-resolved lineage from the accepted predecessor. We define runtime authorization consistency as:
\begin{equation}
\begin{aligned}\small
Consistent(e_i) \equiv\;& LocalAllow(e_i)
\land ValidOrigin(e_i) \\
&\land \bigl((pred_i=\bot \land RootOK(e_i,G)) \\
&\lor (pred_i\neq\bot \land ValidLineage(e_i))\bigr) \\
&\land NoMorePermissive(e_i, B_i^{-}).
\end{aligned}\label{eq:1}
\end{equation}

Here, $LocalAllow$ is the ordinary access-control decision made by the underlying tool or resource layer. $ValidOrigin$ requires each referenced resource to come either from the grant or from a controller-verified anchor. The third conjunct allows either an initial grant-bound event rooted at $\bot$, or a continuation whose predecessor resolves to an accepted record in the same workflow session.
Finally, $NoMorePermissive$ performs the workflow-level check: the continuation must remain within the inherited basis across action, resource, purpose, delegation, and condition constraints.

A workflow is authorization-consistent when every committed authorization-relevant event satisfies this predicate at its pre-commit boundary. Accepted events become part of the recorded lineage, together with the finalized authorization basis used for subsequent checks. Blocked events do not execute and cannot create successor state.

\subsection{Threat Model and Trust Assumptions}
\label{sec:threat-model}

We consider an adversary that can influence the agent's runtime behavior but cannot compromise the controller-mediated enforcement boundary. In particular, the adversary may:
\begin{itemize}
    \item inject malicious instructions through user prompts, retrieved content, tool outputs, intermediate artifacts, or delegated execution contexts;
    \item induce the planner to propose locally plausible continuations with changed resources, shifted purposes, forged predecessor references, amplified delegation, weakened runtime conditions, or unsupported resource identifiers;
    \item provide misleading free-form text, tool-output metadata, or predecessor hints that appear to justify a pending continuation.
\end{itemize}

The adversary's goal is to induce \emph{authorization drift}: a pending workflow step remains locally executable or appears consistent with ordinary tool permissions, while no longer being justified by the authorization basis inherited from accepted predecessor steps. The intended outcome is a continuation that turns an accepted workflow state into authority for actions, data use, or downstream effects that were not authorized for the session.

RAC assumes a controller-mediated enforcement boundary with complete mediation: authorization-relevant tool actions pass through the workflow controller or tool gateway before they become externally effective. The controller-side RAC components are trusted to construct the checked event, verify supporting evidence, maintain accepted authorization state, and enforce pre-commit decisions. RAC relies on deployment-provided authorization inputs, including the session grant, local access-control result, tool manifests, and resource metadata. It preserves consistency with respect to the supplied grant, but does not prove that the grant is least-privileged. When anchors are hash-bound to controller-observed outputs, the hash function is assumed to be collision-resistant.

Planner-facing text and tool self-reports are treated as non-authoritative. They can influence the proposed call or provide diagnostic context, while RAC derives security-critical state only from controller-accepted sources.

\begin{figure*}[t]
\centering
\includegraphics[width=0.9\textwidth]{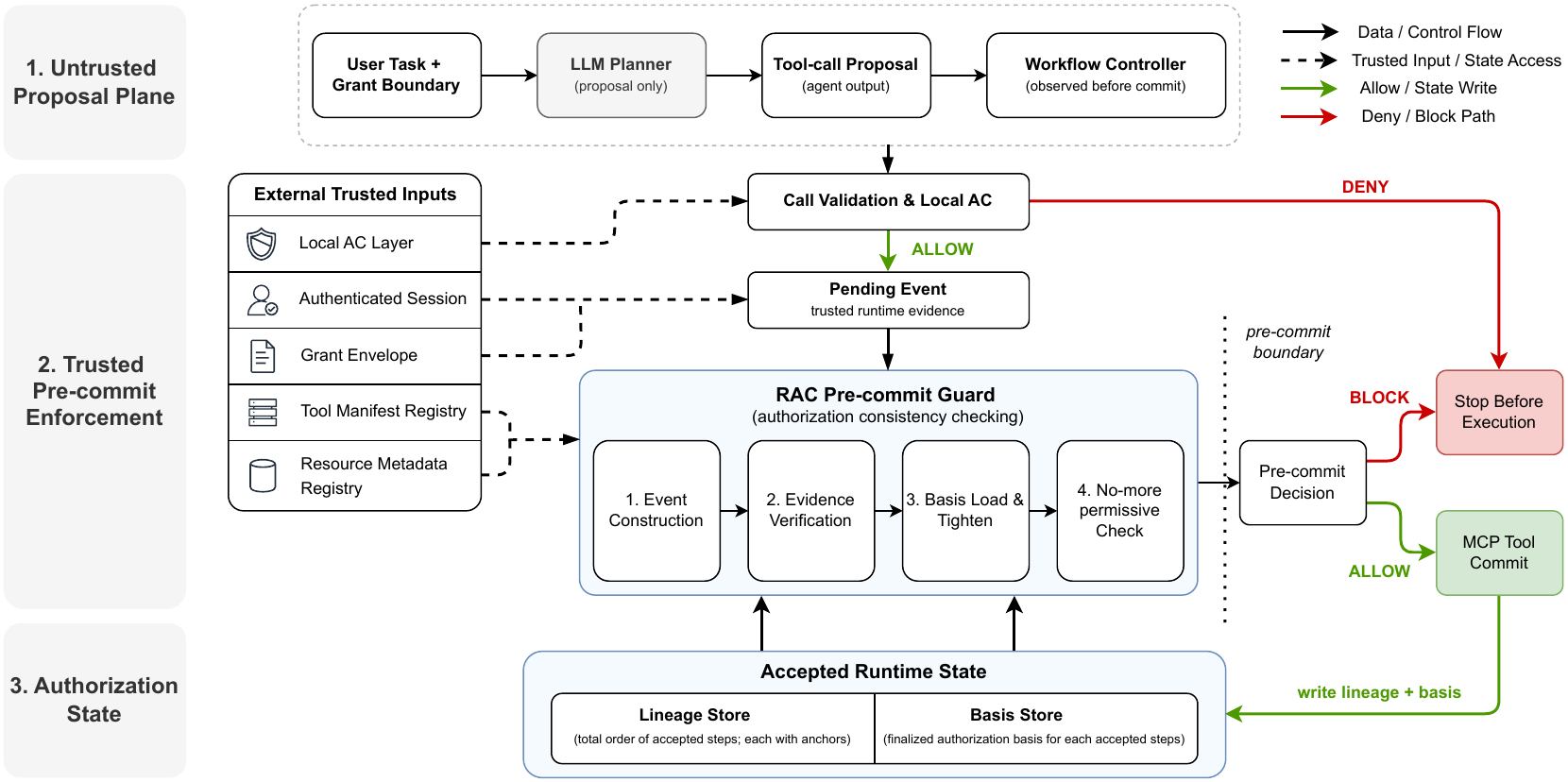}
\caption{System overview of RAC as a pre-commit authorization consistency guard for controller-mediated MCP-style workflows.}\label{fig:overview}
\end{figure*}

\subsection{Scope and Non-goals}
\label{sec:scope-nongoals}

RAC covers authorization-relevant tool actions that pass through the controller before they take external effect. It assumes sufficient structured runtime evidence to construct typed events and resolve lineage; otherwise, the step fails closed. Actions that bypass this boundary, or hidden side effects outside the mediated tool path, are outside its guarantee.

The mechanism assumes that user intent has already been reflected in the grant and focuses on preserving that grant during execution. Least-privilege assessment, grant repair, and semantic completeness of tool outputs are handled by surrounding components such as local access control, policy engines, prompt-injection defenses, and host-level sandboxing.

The current prototype adopts a conservative single-predecessor model. Continuations whose verified inputs resolve to ambiguous or multiple predecessors are blocked. Multi-predecessor joins, alert-level enforcement policies, stronger cross-domain provenance mechanisms, and production-scale drift-frequency measurement are left to future work.

\section{System Overview}
\label{sec:overview}

RAC runs in an instrumented MCP-based workflow controlled by a runtime controller or tool gateway. The controller observes authorization-relevant tool actions before they commit and exposes structured runtime metadata to RAC. Each workflow session is associated with an authenticated subject and a grant envelope that defines the initial authorization boundary. RAC assumes that a trusted authorization layer supplies the initial grant, rather than deriving it from natural-language prompts.

For the motivating workflow, the grant permits Alice to read \code{calendar\_alice} and derive availability for \code{find\_available\_slots}, with no delegation and with \code{contact\_others=false}.

Figure~\ref{fig:overview} shows the runtime placement of RAC. A user task enters a workflow session under the grant boundary above. The planner proposes a tool call, and the workflow controller captures the pending action before execution. The controller first validates the call structure and invokes the local access-control layer for the requested tool or resource operation. If local access control denies the action, the controller rejects the step directly. If the local decision admits the action, the pending step is passed to RAC for workflow-level authorization consistency checking.

RAC checks whether the pending action is still supported by the authorization evidence carried forward from the accepted execution history. The event adapter turns controller-trusted observations into the typed authorization event checked by RAC. The event identifies the pending call within the current grant and records the evidence needed for lineage and resource-origin checks. RAC then loads the inherited basis and decides whether the call remains no more permissive than that basis.

The decision is enforced before tool execution. If RAC returns \textsc{Block}, the controller stops the pending action and no accepted successor state is written. If RAC returns \textsc{Allow}, the controller issues the tool call according to the surrounding workflow policy, and RAC records the admitted step as accepted lineage and finalized basis state for downstream checks.

\section{RAC Design}
\label{sec:design}
RAC enforces runtime authorization consistency by turning accepted workflow history into explicit authorization state. A pending tool call is admitted only when it is supported by the basis carried forward from that history. For each pending call, RAC reconstructs the authorization-relevant event from controller-observed evidence, resolves the accepted predecessor that can support the call, and checks the call against the basis stored for that predecessor. If any step fails, the call is blocked before it reaches the external tool and no successor state is created.

We use the calendar-availability workflow from Section~\ref{sec:motivation} as a running example. The grant permits Alice to read her calendar and derive candidate meeting slots from that calendar output. The authority carried forward from this grant covers calendar-based availability derivation. Email inspection, message drafting, and outbound contact require separate authorization.

\subsection{Trusted Authorization Event Construction}
\label{sec:trusted-event-construction}

RAC begins by separating the planner's proposal from the event that is actually checked. Planner-provided tool choices, arguments, and explanations are treated as proposal data. The event adapter rebuilds the authorization-relevant view of the call from controller-observed runtime evidence, using the session grant, manifest-typed arguments, registered resources, and verified anchors as accepted sources.

RAC performs its check over the constructed authorization event. The action type and resource scope in this event are determined by the tool manifest and verified call arguments, while the purpose field is normalized into a controlled token within the authorized scope. Planner-supplied hints are retained only as diagnostic metadata. The usable predecessor relation is resolved from verified anchors and the lineage store. This prevents the planner from turning its own description of a call into an authorization basis for later steps.

\begin{table*}[t]
\centering
\caption{Source discipline for security-critical fields in RAC events.}
\label{tab:event-sources}
\footnotesize
\setlength{\tabcolsep}{4.2pt}
\renewcommand{\arraystretch}{1.08}
\begin{tabular*}{\textwidth}{@{\extracolsep{\fill}}p{0.18\textwidth}p{0.35\textwidth}p{0.38\textwidth}@{}}
\toprule
Field & Source accepted by RAC & Not accepted as sole evidence \\
\midrule
Subject & Authenticated session context & LLM explanation or tool-output text \\
Tool & Workflow controller boundary & Natural-language tool description \\
Action & Tool manifest and action profile & Agent-declared action label \\
Resource & Manifest-typed tool arguments and registered resource identifiers & Free-form resource mentions \\
Resource origin & Grant scope or controller-verified predecessor anchors & Resource lineage mentioned only in text \\
Purpose & Grant-bound canonical purpose token & Free-form purpose phrase before normalization \\
Delegation & Grant envelope and controller state & Agent self-declared delegation claim \\
Conditions & Grant constraints and runtime metadata & Prompt-level condition statements \\
Input anchors & Anchors verified by the controller & Anchors reported only by tool output \\
Predecessor & Lineage store resolved from verified anchors & Agent-declared predecessor hints \\
\bottomrule
\end{tabular*}
\end{table*}

Table~\ref{tab:event-sources} summarizes this source discipline. Every security-critical field used by RAC must be grounded in an accepted runtime source. Free-form text can assist normalization or diagnostics, but it is not used as standalone authorization evidence. If a required field cannot be grounded, event construction fails and the pending step is rejected.

In the running example, the pending \code{read\_calendar} call becomes a valid authorization event only after the controller grounds the call in Alice's session, the manifest-derived operation, the validated calendar resource, and the grant-bound purpose. If required evidence is missing or the requested call is not covered by the authorized profile, the adapter cannot construct a valid event and RAC rejects the pending step.

\subsection{Anchor and Resource-Origin Verification}
\label{sec:anchor-resource-verification}
RAC uses anchors to decide when an accepted output can support later continuation. The controller first binds the claimed anchor to the observed output, the producer step, and the schema admitted by the tool manifest. Once this binding is established, resource-origin verification can treat the anchor as authorization-relevant evidence. A pending step can rely on a resource only when that resource is covered by the grant or traced to a verified anchor. This keeps generated references from being promoted into authorization support.

\textbf{Anchor verification.}
When a tool output is intended to support later continuation, RAC first treats the corresponding anchor claim as raw evidence. The controller verifies the claim against the output it actually observes at the producing step and the resources derivable at the tool boundary. Let $y_i$ denote the concrete output observed by the controller for step $i$, and let $ObsRes_i$ denote the resources derived from checked arguments, runtime access records, and manifest-declared mappings. A raw anchor claim $a_i^{raw}$ becomes a verified anchor only when:
\begin{equation}
\small
\begin{aligned}
VerifyAnchor(a_i^{raw}) \equiv\;&
producer(a_i^{raw}) = i \\
&\wedge\; h_i^{claim}=H(\mathsf{canon}(y_i)) \\
&\wedge\; Res_i^{claim}\subseteq ObsRes_i \\
&\wedge\; schema(a_i^{raw})\in\mathcal{S}_{tool_i}.
\end{aligned}
\end{equation}

Here, $h_i^{claim}$ is the claimed content hash, $Res_i^{claim}$ is the claimed resource set, and $\mathcal{S}_{tool_i}$ is the set of anchor schemas admitted by the tool manifest. In the running example, the calendar-read output yields a verified anchor only when the claim matches the controller-side evidence for the producer, output hash, resource set, and schema. The claimed resource must be contained in $\{\code{calendar\_alice}\}$, and the schema must be admitted by the manifest, such as \code{availability}/\code{v1}. In our prototype, $\mathsf{canon}(y_1)$ is computed through deterministic controller-side serialization, using raw bytes for filesystem artifacts and stable serialization for structured outputs. Only verified anchors can be attached to lineage records or used in downstream resource-origin checks.

\textbf{Resource-origin verification.}
RAC separates resource existence from authorization origin. A registry entry establishes existence, while workflow authority must come from the grant or verified lineage evidence. For each resource identifier referenced by a pending event $e_i$, RAC accepts its origin only when the resource is covered by the workflow grant or carried by a verified input anchor:

\begin{equation}
\small
\begin{aligned}
ValidOrigin(e_i) &\triangleq
\forall res \in Res(e_i):
 res \in Scope_{res}(G_i) \\
& \vee \exists a \in I_i^{ver}: res \in Resources(a).
\end{aligned}
\end{equation}

Here, \(Res(e_i)\) denotes the resources referenced by the pending event, \(G_i\) is the session grant envelope, \(Scope_{res}(G_i)\) is the resource scope defined by the grant, \(I_i^{ver}\) denotes the verified input anchors attached to \(e_i\), and \(Resources(a)\) returns the controller-verified resources carried by anchor \(a\). If any referenced resource fails this predicate, RAC emits a resource-origin violation and blocks the step before commit.

For example, a generated artifact can mention \code{email\_thread\_conflict}, after which the planner proposes a \code{search\_email} call over that thread. Even if the registry confirms that the thread exists, RAC admits the call only if the thread is included in the grant envelope or appears in a verified input anchor.

\subsection{Controller-Resolved Causal Lineage}
\label{sec:causal-lineage}

A continuation inherits authorization from an accepted step resolved by the controller. RAC derives this predecessor from verified input anchors and accepted lineage records. Lineage is therefore checked as an authorization dependency: when a pending call consumes an artifact, RAC must identify the accepted step that produced it in the same workflow session, with matching anchor evidence.

The controller maintains an append-only lineage store for accepted steps. Each record binds the accepted step to its session, verified data-flow evidence, predecessor relation, and finalized basis:

\begin{equation}
\small
L_i=\langle sid_i,seq_i,pred_i,I_i^{ver},O_i^{ver},op_i,bid_i\rangle .
\end{equation}

Here, $sid_i$ and $seq_i$ identify the workflow session and controller-assigned order, $pred_i$ links to the accepted predecessor, $I_i^{ver}$ and $O_i^{ver}$ record verified input and output anchors, $op_i$ summarizes the controller-observed operation, and $bid_i$ points to the finalized authorization basis. The record is hash-linked as:

\begin{equation}
\small
hash_i = H(\mathsf{enc}(sid_i, seq_i, pred_i, I_i^{ver}, O_i^{ver}, op_i, bid_i)).
\end{equation}

This binding ties the accepted sequence, verified data flow, and successor basis to the same controller-observed step. A later continuation cannot reinterpret an output under a different predecessor or load a basis from an unrelated accepted event.

For a pending continuation event, lineage resolution starts from its verified input anchors. RAC queries the lineage store for the accepted producer of each input anchor and checks that the resolved producer belongs to the same session. In the current prototype, lineage validity requires all verified input anchors to resolve to the same accepted predecessor in the same session. Each input anchor must match a recorded output anchor for that predecessor, including the anchor identifier and content hash. Failed resolution, ambiguous producers, hash mismatches, or session mismatches produce a lineage violation before the pending step commits.

The resolved predecessor is then used to load the finalized authorization basis for propagation. For an initial grant-bound event with no predecessor, RAC initializes the basis directly from the grant envelope and uses the null predecessor $\bot$. Predecessor hints carried by the event are advisory metadata used for candidate lookup or diagnostics. Accepted lineage is established through verified anchors and controller-recorded lineage entries. The prototype therefore rejects ambiguous or multi-predecessor continuations by default. Supporting benign multi-source continuations requires an explicit merge policy, as discussed in Section~\ref{sec:limitations}.

Under this design, predecessor selection is determined by controller-verified evidence. A downstream tool call can inherit authorization only through accepted, controller-observed execution history. Fake anchors, outputs that were never accepted into lineage, cross-session references, and anchors whose hashes do not match recorded outputs fail resolution before the pending action commits.

\subsection{Authorization Basis Propagation and Tightening}
\label{sec:basis-propagation}

After resolving the predecessor, RAC retrieves the authorization basis associated with the accepted step, which captures the authority that remains available for continuation. This basis comes from previously admitted execution state and is updated only after the pending event passes the pre-commit check. RAC then checks the pending event against this inherited basis before commit, updating the basis for subsequent steps only if the event is admitted.

For the first authorization-relevant step of a workflow, the basis is initialized from the grant envelope. For a continuation step, RAC loads the finalized basis stored for the resolved predecessor:
\begin{equation}
\small
B_i^- = LoadBasis(pred_i).
\end{equation}
The pending event is checked against this inherited basis and, if admitted, used to produce a tightened successor basis:
\begin{equation}
\small
B_i= Tighten(B_i^-, Obs(e_i)).
\end{equation}
Here, $Obs(e_i)$ denotes the authorization-relevant observation extracted from the pending event.

For actions, RAC uses an explicit coverage relation over canonical action labels. We write $a_1 \preceq_{\mathsf{act}} a_2$ when action $a_1$ is no more permissive than action $a_2$ and can therefore be covered by a basis that allows $a_2$. The relation is specified by the deployment manifest and grant profile, represented as an acyclic action graph in the prototype, and tested as part of the authorization configuration. Labels are treated as incomparable unless a coverage relation is explicitly configured. For an action set $S$, let $Down(S)=\{a \in \mathcal{A}\mid \exists a' \in S: a \preceq_{\mathsf{act}} a'\}$ denote the downstream-closed action set covered by $S$, where $\mathcal{A}$ is the set of canonical action labels. We use $Allow_{\mathsf{act}}(B)$ to denote the action labels that remain admissible under basis $B$ after applying this relation.

The prototype realizes this relation through deployment-validated manifest mappings. Authorization-relevant tool operations are checked only after they map to canonical action labels. The coverage graph is audited against the grant profile, kept acyclic, and interpreted fail-closed for missing labels. Auxiliary tooling can suggest labels or edges, while RAC makes decisions from the audited configuration.

After an event is admitted, RAC derives a successor basis from the predecessor basis and the applicable continuation profile. The profile comes from the manifest, the grant profile, or a verified output anchor, and determines the authority available for later continuation. When no narrower profile is available for a dimension, RAC preserves the inherited basis for that dimension. The successor basis is computed as follows:
\begin{equation}
\small
\begin{aligned}
Res(B_i) &= Res(B_i^-)\cap Cont_{\mathsf{res}}(e_i), \\
Purp(B_i) &= Purp(B_i^-)\cap Cont_{\mathsf{purp}}(e_i), \\
Allow_{\mathsf{act}}(B_i) &= Allow_{\mathsf{act}}(B_i^-)\cap Cont_{\mathsf{act}}(e_i), \\
Cond(B_i) &= TightenCond(Cond(B_i^-), cond(e_i)).
\end{aligned}
\end{equation}

Here, $Cont_{\mathsf{res}}$, $Cont_{\mathsf{purp}}$, and $Cont_{\mathsf{act}}$ denote the continuation profiles used for the admitted event. If any dimension becomes empty or incompatible, RAC blocks the event before commit and does not create a successor basis. In the running example, the verified calendar output supports availability derivation from \code{calendar\_alice}. Email-thread access and external-contact actions remain outside the successor basis, so searching related messages or sending a rescheduling email requires separate authorization.

\begin{table*}[t]
\centering
\caption{Rule families for enforcing runtime authorization consistency.}
\label{tab:precommit-rules}
\footnotesize
\setlength{\tabcolsep}{4.0pt}
\renewcommand{\arraystretch}{1.06}
\begin{tabular*}{\textwidth}{@{\extracolsep{\fill}}p{0.1\textwidth}p{0.36\textwidth}p{0.44\textwidth}@{}}
\toprule
Rule & Check & Violation evidence \\
\midrule
LineageOK & Resolve one accepted predecessor via verified anchors. & Missing, cross-session, ambiguous, or hash-mismatched predecessor. \\
OriginOK & Trace each resource to the grant or verified anchors. & Resource exists but has no authorized workflow origin. \\
SubjectOK & Match subject and effective subject to the inherited basis. & Subject substitution or effective-subject mismatch. \\
ActionOK & Require action coverage by the inherited basis. & Action escalation or uncovered canonical action. \\
ResourceOK & Keep resources within the inherited scope. & Resource expansion beyond the carried basis. \\
PurposeOK & Keep purpose within the inherited scope. & Purpose drift across workflow continuation. \\
DelegationOK & Preserve grant-bound delegation constraints. & Unauthorized delegatee, redelegation, or depth violation. \\
ConditionOK & Preserve inherited runtime constraints. & Time, tenant, environment, channel, or label violation. \\
\bottomrule
\end{tabular*}
\end{table*}

\subsection{Pre-commit Consistency Checking}
\label{sec:precommit-checking}

RAC evaluates each pending event after local access control admits the tool-call proposal and before the controller commits the corresponding action. At this point, the checker has a typed event $e_i$ and the inherited basis $B_i^-$ loaded from the resolved predecessor. RAC therefore focuses on cross-step consistency and operates alongside local access control.

Table~\ref{tab:precommit-rules} summarizes the rule families that instantiate the no-more-permissive relation.

The action check uses the configured coverage relation introduced in Section~\ref{sec:basis-propagation}. This relation is part of the deployment's manifest and grant specification, and RAC uses the configured relation as the sole source of action coverage. A pending event passes the action check when its requested action belongs to the action set still allowed by the inherited basis:

\begin{equation}
\small
ActionOK(e_i,B_i^-)\equiv act(e_i)\in Allow_{\mathsf{act}}(B_i^-).
\end{equation}

The other rule families apply the same containment idea to their own fields, with the required evidence listed in Table~\ref{tab:precommit-rules}. If a rule fails, RAC records what failed, which field was involved, and which evidence supported the failure. Section~\ref{sec:decision-state} then uses these recorded violations to make the final decision.

\subsection{Decision and Persistent State}
\label{sec:decision-state}

The rule evaluator returns a violation set $V_i$ for the pending event. RAC maps this set to a pre-commit control decision:
\begin{equation}
Decision(e_i) =
\begin{cases}
\textsc{Block}, & V_i \neq \emptyset,\\
\textsc{Allow}, & V_i = \emptyset.
\end{cases}
\end{equation}

The decision determines whether the event is admitted into RAC's runtime authorization state. If RAC returns \textsc{Block}, the controller stops the pending action before tool execution and RAC writes no lineage or successor basis for that event. The accepted lineage available to later steps therefore contains no predecessor, anchor, or authority derived from the rejected event.

If RAC returns \textsc{Allow}, the controller issues the tool call according to the surrounding workflow policy. RAC then appends the verified lineage entry and stores the finalized successor basis produced by tightening. Later events must resolve their predecessor through this accepted lineage and load the corresponding basis incrementally. This state update lets later checks inherit authorization only from accepted workflow steps.

\section{Security Argument}
\label{sec:security-argument}

We now show that, under the trust assumptions in Section~\ref{sec:threat-model}, RAC preserves the runtime authorization consistency predicate in Eq.~\ref{eq:1}. The argument is a safety argument over authorization-relevant events that pass through the monitored controller boundary. RAC's guarantee is that every committed event remains supported by the authorization basis inherited from accepted workflow history. Appendix~\ref{app:proof-details} gives lemma-level proof details.

\noindent\textbf{Theorem 1 (Consistency of committed events).} Assume that authorization-relevant tool actions pass through the controller-mediated boundary, the controller-side RAC components construct event fields according to Table~I, verified anchors are hash-bound to controller-observed outputs, and the lineage and basis stores are not compromised. For every event $e_i$ that RAC allows and commits, $Consistent(e_i)$ holds at the pre-commit boundary. Events blocked by RAC do not create successor lineage or basis state.

\noindent\textit{Proof sketch.} Consider a pending event $e_i$ that receives an \textsc{Allow} decision. The event adapter derives security-critical fields only from accepted runtime sources, so RAC checks a controller-grounded event whose fields are not established by planner declarations. The local access-control result supplies $LocalAllow(e_i)$. Anchor and resource-origin verification establish that every referenced resource is either grant-bound or carried by a verified predecessor anchor, which gives $ValidOrigin(e_i)$. For an initial grant-bound event, the $\bot$ predecessor and initialization from $G$ satisfy the root case in Eq.~(\ref{eq:1}); otherwise, lineage resolution accepts only same-session predecessors whose verified input anchors match recorded output anchors in the lineage store, satisfying the continuation case. The rule evaluator returns \textsc{Allow} only when no rule in Table~II reports a violation; therefore the requested continuation remains within the inherited basis, establishing $NoMorePermissive(e_i,B_i^-)$.

After an allowed event commits, RAC appends its verified lineage record and stores the tightened successor basis. The tightening operator preserves or narrows the inherited basis according to the configured continuation profile, using only accepted runtime evidence. If any required check fails, RAC returns \textsc{Block} and the controller stops before issuing the external call or writing successor state. By induction over the accepted sequence of workflow steps, every committed authorization-relevant event satisfies Eq.~(\ref{eq:1}), while later continuations can draw authorization evidence only from accepted events.

\section{Prototype and Implementation}\label{sec:implementation}

We implemented RAC as a Python runtime guard at the host-side tool-call boundary. The guard receives pending authorization-relevant tool actions, applies the checks defined in Section~\ref{sec:design}, and returns an allow/block decision to the workflow controller before the call reaches the server. This placement lets the client observe proposed tool calls before execution, without requiring changes to the MCP protocol or server-side tool implementations.

\subsection{Prototype Architecture}

The implementation follows the design pipeline, covering event adaptation, evidence verification, lineage resolution, basis management, consistency checking, and decision mapping. For evaluation, the prototype uses in-memory lineage and basis stores. The lineage store keeps the accepted execution history and the evidence needed for predecessor resolution, while the basis store keeps the finalized authorization basis after each admitted step. These stores keep each decision incremental, depending only on the pending event, the resolved predecessor, and the predecessor basis. Durable storage, distributed tracing, or tamper-evident logging can replace the in-memory stores in production deployments without changing the checking semantics.

\subsection{MCP Integration and Reproducible Artifacts}

The artifact package contains controlled traces, ablation scripts, latency scripts, and fixed filesystem replay workflows for reproducing the evaluation. For live-boundary validation, we implement a real MCP filesystem case study using the \texttt{@modelcontextprotocol/server-filesystem} server over stdio. A guarded client invokes RAC before issuing each filesystem operation. If RAC returns \textsc{Block}, the client stops the call and records that no server call was issued; if RAC returns \textsc{Allow}, the call proceeds and the accepted state is recorded for downstream checks. This setup evaluates pre-call enforcement at a live MCP boundary, with no changes to the MCP protocol or the filesystem server.

The artifact set also includes fixed replay corpora for the planner-generated and MCP filesystem studies. For RQ2, the LLM planner is used only offline to generate candidate planner-style workflows; the generated plans are normalized into deterministic traces, and the reported observable subset is derived from explicit tool-call facts. For RQ3, the deterministic filesystem replay corpus combines hand-authored planner-style plans with offline LLM-generated plans normalized into fixed JSON workflows. During replay, RAC consumes only these fixed workflows and does not invoke the LLM in the enforcement loop. The artifact package records the normalized plans, subset construction metadata, and validation scripts so that guarded and unguarded executions can be reproduced under the same workflow inputs.

\section{Evaluation}
\label{sec:evaluation}

\subsection{Experimental Setup and Artifacts}
\label{sec:evaluation-setup}

We evaluate RAC along four questions. RQ1 uses TraceBench to test whether RAC enforces the authorization-consistency specification and how much of the same specification workload is covered by simpler enforcement strategies. RQ2 complements TraceBench with planner-generated workflows and evaluates a high-confidence observable subset where the expected decision follows from explicit tool-call facts. RQ3 uses a real MCP filesystem server over stdio to test whether unsafe continuations are stopped before server execution. RQ4 measures the component-level latency of RAC's checking path.

Oracle labels are generated from the specification predicates before running the RAC implementation. For each workflow, the benchmark records the initial grant, the accepted prefix, the pending continuation, and the evidence available at the pre-commit boundary. The expected decision is derived from whether the pending continuation violates the corresponding consistency predicate. This separates labeling from implementation output: RAC and the baselines are evaluated on the same workflow evidence and oracle decisions.

TraceBench is derived from the threat model in Section~\ref{sec:threat-model}. Each workflow starts from an otherwise admissible task context and introduces a locally plausible continuation influenced by one of the adversary capabilities: expanding the requested action or resource scope, changing purpose or runtime conditions, amplifying delegation, injecting unsupported evidence, or manipulating predecessor support. The resulting continuation can still be executable under ordinary tool permissions, while violating the authorization basis inherited from accepted workflow history. Each case is paired with an oracle decision and the predicate that determines the expected outcome.

The TraceBench suite contains 1,248 workflows, including 240 oracle-ALLOW workflows and 1,008 oracle-BLOCK workflows. It covers benign continuations, single-drift objectives, and composite drift cases over varied actions, resources, purposes, runtime conditions, and evidence patterns. The planner-generated replay corpus contains 300 workflows generated by an external instruction-tuned LLM planner. The planner is used only for blind plan generation: it receives natural-language tasks, tool descriptions, and contextual resources, without access to RAC, oracle labels, or drift categories. These plans are normalized into fixed replay traces before evaluation, and the LLM is not invoked in the enforcement loop.

\begin{figure*}[t]
\centering
\makebox[\textwidth][c]{%
\begin{subfigure}[t]{0.4\textwidth}
\centering
\includegraphics[width=\linewidth]{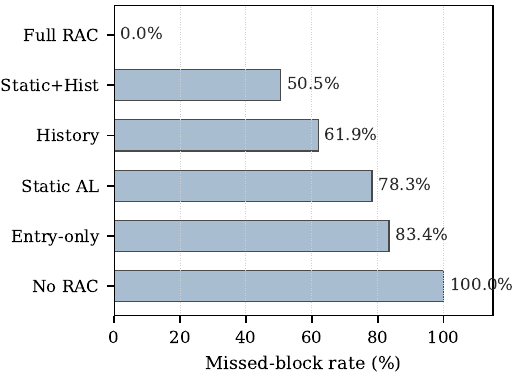}
\caption{Overall missed-block rate}
\label{fig:tracebench-baselines}
\end{subfigure}
\hspace{0.06\textwidth}
\begin{subfigure}[t]{0.4\textwidth}
\centering
\includegraphics[width=\linewidth]{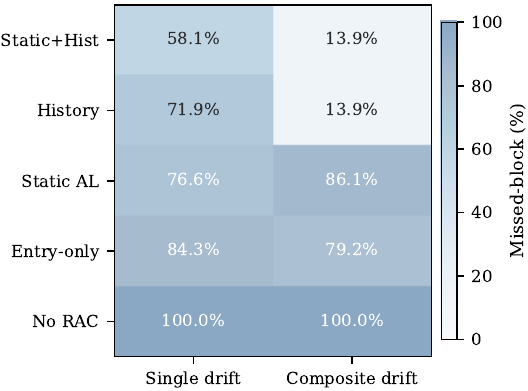}
\caption{Baseline blind spots}
\label{fig:tracebench-blindspots}
\end{subfigure}%
}
\caption{TraceBench baseline comparison on the controlled specification workload.}
\label{fig:tracebench-baseline-comparison}
\end{figure*}

\subsection{RQ1: TraceBench Specification Coverage}
\label{sec:rq1-tracebench}

RQ1 evaluates RAC on TraceBench, a controlled workload designed to exercise the authorization-drift patterns instantiated by our rule families. The experiment tests whether the implementation enforces the modeled authorization-consistency predicate at the pre-commit boundary, and how much missed-block reduction comes from checking workflow evidence beyond entry-time or per-tool authorization. We compare Full RAC with five baselines: No RAC, Entry-only checking, a static tool-allowlist heuristic, a history-aware resource-scope checker, and a combined Static+History baseline. Entry-only checking validates only the initial grant boundary and does not check continuation evidence. The static allowlist admits calls to approved tools without reasoning about inherited workflow evidence. The history-aware baseline admits a step only when its resources are contained in the initial grant or in resources accumulated from previously accepted steps. Static+History combines the static tool filter with this resource-history check.

Figure~\ref{fig:tracebench-baseline-comparison} reports missed-block rates on TraceBench. Full RAC has no missed-block case on this workload. In contrast, No RAC misses all oracle-BLOCK workflows, Entry-only checking misses 841/1,008 cases, the static tool allowlist misses 789/1,008 cases, and the history-aware resource-scope checker misses 624/1,008 cases. Static+History is the strongest baseline, but still misses 509/1,008 oracle-BLOCK workflows. These results show that local tool filtering and resource-history tracking cover some visible expansions, but leave many workflow-level violations outside their modeled state.

Figure~\ref{fig:tracebench-blindspots} breaks down the baseline misses by scenario type. Tool allowlists mainly help when a continuation invokes a disallowed interface. History-aware checks help when a continuation introduces an unseen resource. They are less effective when the decision depends on purpose preservation, runtime conditions, delegation constraints, output anchors, or lineage validity. Composite cases are not uniformly harder for every baseline: when a composite continuation includes a visible resource or action expansion, a simple baseline may block it. The remaining misses arise when the violation is carried by evidence or inherited authority that the baseline does not represent.

Full RAC covers these cases because it reconstructs the pending call as a trusted authorization event and compares it with the authorization basis inherited through verified lineage. These results show why authorization drift cannot be reduced to tool availability or resource-history checks alone. Runtime consistency requires preserving the authority carried forward from accepted workflow steps and rejecting continuations that are more permissive than that inherited basis.

\subsection{RQ2: Planner-Generated Observable Drift Replay}
\label{sec:rq2-llm-observable}

RQ2 complements TraceBench with planner-generated workflows. We generated 300 planner-style workflows using an external instruction-tuned LLM in a blind proactive-planning setting. The model received task descriptions, available tools, and contextual resources, without access to RAC, oracle labels, or drift categories. The generated plans were normalized into deterministic replay traces. Together, these traces test whether RAC's enforcement behavior observed on TraceBench persists on model-produced workflow continuations, including cases with observable authorization-drift opportunities.

To avoid relying on ambiguous natural-language interpretation, we report only a high-confidence observable subset. A decision instance is included when the expected decision follows from explicit tool-call facts, including grant-bound reads, summaries over verified prior reads, unauthorized resource accesses, external-effect actions, external-purpose arguments, or unverified anchors. We exclude decisions whose labels depend only on free-form rationale or ambiguous natural-language anchors. The resulting subset contains 898 decision instances across 300 workflows: 605 oracle-ALLOW and 293 oracle-BLOCK.

\begin{table}[t]
\centering
\caption{Missed-block analysis on planner-generated observable drift decisions. Metrics are computed over oracle-BLOCK decisions reached by each variant.}
\label{tab:llm-observable-subset}
\small
\setlength{\tabcolsep}{3.0pt}
\renewcommand{\arraystretch}{1.08}
\begin{tabular}{@{}lrrrr@{}}
\toprule
Variant & Reached BLOCK & Blocked & Missed & Recall \\
\midrule
Full RAC & 223 & 207 & 16 & 0.928 \\
Static+Hist. & 237 & 163 & 74 & 0.688 \\
History-aware & 243 & 163 & 80 & 0.671 \\
Static allowlist & 237 & 80 & 157 & 0.338 \\
\bottomrule
\end{tabular}
\end{table}

Table~\ref{tab:llm-observable-subset} focuses on missed blocks, the security-relevant failure mode in this study. Full RAC blocks 207 of 223 reached oracle-BLOCK decisions, achieving a block recall of 92.8\% and a missed-block rate of 7.2\%. Static+History and History-aware reach 68.8\% and 67.1\% block recall, respectively, while the static tool allowlist reaches 33.8\%. As a lower-bound control, No RAC reaches all 293 oracle-BLOCK decisions and blocks none. On reached oracle-ALLOW decisions in this subset, the evaluated variants produced no over-blocks; we therefore focus the table on missed blocks.

The per-drift breakdown explains these differences. Static tool allowlists mainly cover action escalation. History-aware and Static+History cover visible resource expansion and many action cases, but do not reason about purpose or lineage/origin evidence. Full RAC covers the reached action, resource, and purpose-drift decisions in this subset. The remaining misses are concentrated in lineage/origin cases where generated plans refer to prior notes or derived anchors with incomplete trust-status evidence in the normalized replay. We treat these cases as an evidence-visibility limitation of the current replay adapter.

\subsection{RQ3: Real MCP Filesystem Enforcement}
\label{sec:rq3-mcp-filesystem}

We next evaluate whether RAC can block unsafe operations before they reach a real MCP server. We use the MCP filesystem server over stdio and compare guarded and unguarded execution. Full RAC uses a guarded client that invokes RAC before each filesystem operation, while No RAC calls the server directly. The server is launched from the \texttt{@modelcontextprotocol/server-filesystem} package over a sandboxed filesystem root, and the client records whether each call reaches the server before any filesystem side effect occurs.

The evaluation includes a direct filesystem scenario suite and a deterministic planner replay corpus. The replay corpus contains six hand-authored planner-style plans and eight plans generated offline by an LLM. We normalize all plans into fixed JSON workflows before evaluation, mapping tool names, arguments, and JSON types into the replay schema while preserving oracle labels and authorization evidence. Replay is deterministic and uses the same workflow inputs for guarded and unguarded execution. RQ3 therefore measures whether RAC enforces authorization-continuity checks at the live MCP-style tool boundary under fixed workflow inputs.

\begin{table}[t]
\centering
\caption{Real MCP filesystem enforcement results.}
\label{tab:mcp-filesystem}
\footnotesize
\setlength{\tabcolsep}{3.0pt}
\renewcommand{\arraystretch}{1.12}
\begin{tabular}{@{}p{0.24\columnwidth}rccccc@{}}
\toprule
Setting & Steps & \multicolumn{2}{c}{Full RAC} & \multicolumn{2}{c}{No RAC} & Unsafe \\
\cmidrule(lr){3-4}\cmidrule(lr){5-6}
 & & Match & Stopped & Match & Stopped & Side effects \\
\midrule
Direct scenarios & 3 & 3/3 & 2 & 1/3 & 0 & 2 $\rightarrow$ 0 \\
Planner corpus & 29 & 29/29 & 9 & 20/29 & 0 & 5 $\rightarrow$ 0 \\
\bottomrule
\end{tabular}

\vspace{2pt}
\begin{minipage}{0.96\columnwidth}
\footnotesize
\emph{Note:} Unsafe side effects are shown as No RAC $\rightarrow$ Full RAC.
Stopped denotes calls blocked before reaching the MCP server.
\end{minipage}
\end{table}

One representative blocked workflow starts from a grant that permits reading an admitted project file for summarization. After the read, the replayed planner proposes writing the derived content to an outgoing shared path. Local filesystem permission admits the write, so No RAC materializes the side effect. RAC blocks the step before the server call because the requested write action and output path are not covered by the inherited basis or any verified resource-origin anchor.

Across the direct scenarios and planner corpus, Full RAC matches all expected decisions, stops unsafe calls before execution, and prevents all unsafe filesystem side effects. No RAC executes the locally permitted calls and materializes five unsafe side effects in the planner corpus. These results show that RAC stops rejected continuations before the MCP filesystem server observes the call.

\subsection{RQ4: Runtime Overhead}
\label{sec:rq4-overhead}

Finally, we measure the runtime overhead of RAC during full TraceBench replay. The benchmark replays the 1,248-workflow suite under Full RAC for 20 iterations, producing 24,960 trace replays and 59,120 step-level latency records. For each replay, we record component-level latency for output-anchor prechecking, lineage resolution, the pre-commit consistency check, total step latency, and total trace replay latency. This setup measures the checking path exercised by the TraceBench workload used in RQ1.

\begin{figure}[t]
\centering
\includegraphics[width=0.85\columnwidth]{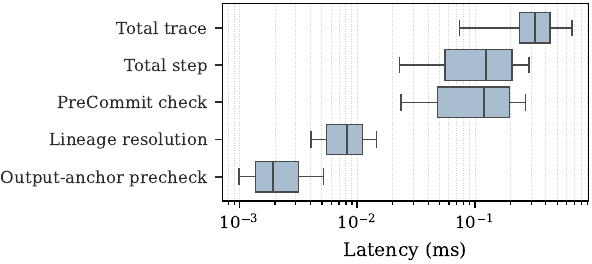}
\caption{RAC checking-path latency on full TraceBench replay. Boxes show IQR; whiskers show the 5th--95th percentiles.}
\label{fig:overhead}
\end{figure}

Figure~\ref{fig:overhead} shows the latency distribution across RAC checking stages. The dominant component is the pre-commit consistency check, while output-anchor prechecking and lineage resolution remain comparatively small. Across the full-suite replay, RAC keeps step-level checking latency in the sub-millisecond range: the p50 total-step latency is 0.118 ms, the p95 latency is 0.275 ms, and the p99 latency is 0.377 ms. At the trace level, the p50 replay latency is 0.308 ms, with p95 and p99 latencies of 0.652 ms and 0.872 ms, respectively.

This measurement isolates RAC's in-memory checking path. Deployment integrations with durable stores, remote policy services, distributed tracing, or logging add costs outside the core consistency check.

\section{Discussion}
\label{sec:discussion}

\subsection{Deployment Considerations}
\label{sec:deployment-considerations}

A practical deployment should place RAC before tool calls take external effect. Filesystem, subprocess, and network capabilities should either pass through the same controller or be confined by host policy. This placement is the main operational requirement for RAC's guarantee.

RAC complements existing guardrails by checking whether a proposed continuation is still supported by the inherited authorization basis after it has been proposed and locally admitted. The extracted event and violation set can also support deployment-specific responses, such as alerting, human approval, or quarantine when evidence is incomplete.

\subsection{Limitations}
\label{sec:limitations}

RAC provides a scoped safety guarantee for authorization-relevant actions that pass through the monitored controller boundary. Hidden side effects and direct host capabilities fall outside this boundary and require mediation, monitored-tool wrapping, sandboxing, or egress controls. This is consistent with the threat model in Section~\ref{sec:problem-threat} and the safety argument in Section~\ref{sec:security-argument}: RAC preserves authorization consistency for committed events it can observe and mediate.

RAC also depends on deployment-specific evidence handling and configuration. Verified anchors require stable canonicalization before hashes are compared, and the action coverage relation must be defined over auditable manifest mappings and canonical labels. Incorrect mappings can weaken enforcement, so missing labels fail closed and the checked relation remains explicit. Production deployments can replace the prototype's in-memory lineage and basis stores with durable logs, signed trace context, or stronger artifact provenance without changing the core checking semantics.

Finally, our evaluation should be read as a validation of RAC's authorization-consistency semantics under controlled workloads. TraceBench covers the drift patterns studied in this paper, the observable LLM-plan replay tests fixed planner-generated traces with explicit tool-call evidence, and the real MCP filesystem study demonstrates pre-call blocking at a live boundary. These experiments evaluate semantic enforcement rather than real-world drift prevalence. The current prototype adopts a conservative single-predecessor discipline, which can reject benign joins such as summaries that combine independently authorized reports. Supporting such joins requires an explicit merge operator that preserves compatibility across predecessor bases and keeps resource origins provenance-separated.

\subsection{Ethical Considerations}

Our evaluation uses synthetic TraceBench workflows, normalized offline planner-generated traces, and a sandboxed MCP filesystem root. It does not involve human subjects, private user data, production accounts, or vulnerability testing against third-party systems. The planner-generated workflows are used only as fixed replay inputs after normalization, and no LLM is invoked in RAC's enforcement loop or in the reported decision procedure.

The filesystem study is confined to a temporary test root. Unsafe operations are used only to compare guarded and unguarded replay, and any side effects are limited to controlled test files inside that root. The main deployment risk is over-interpreting the guarantee: RAC covers authorization-relevant actions mediated by the controller boundary, while direct host capabilities still require sandboxing, monitored-tool wrapping, or egress controls.

\section{Related Work}
\label{sec:related-work}

\subsection{LLM Agents and Tool Security}

Recent work on LLM-integrated applications has shown that model behavior can be manipulated through direct or indirect prompt injection, retrieved content, and tool-facing instructions \cite{perez2022ignore,greshake2023not,liu2023prompt}. A parallel line of work evaluates tool-using agents under unsafe action selection, high-stakes tool use, and adversarial or untrusted environments \cite{ruan2024identifying,debenedetti2024agentdojo,zhan2024injecagent}. These studies motivate a trust-boundary lesson for tool-integrated agents: model-facing text should not by itself determine security decisions. RAC applies this experience to workflow authorization, focusing on the supporting relationship between pending continuation operations and accepted execution history. This focus differs from defenses designed to prevent rapid tampering or the selection of insecure tools.

\subsection{Workflow Authorization and Delegation}

Capability-based systems such as Macaroons support attenuation and contextual caveats for decentralized authorization~\cite{birgisson2014macaroons}. In Web authorization, OAuth scopes define bounded access tokens for third-party applications and APIs~\cite{hardt2012oauth}. Workflow authorization and delegation models study constraints, authorization propagation, separation of duties, and conditional delegation in structured processes~\cite{bertino1999specification,li2003delegation,atluri2005supporting}. RAC uses grants, scopes, and delegation constraints as the initial authorization boundary, then checks how that boundary is preserved as an agent toolchain expands at runtime. RAC differs operationally: it checks controller-observed continuation steps online and updates the inherited basis during execution, so the workflow need not be fixed in advance.

\subsection{Runtime Enforcement and Provenance}

Runtime enforcement and reference-monitor designs mediate execution to ensure that security policies hold before protected actions take effect~\cite{schneider2000enforceable,erlingsson2000irm}. Information-flow systems such as Flume and HiStar track labels or decentralized information-flow policies across operating-system abstractions~\cite{krohn2007information,zeldovich2006making}. Provenance systems capture system-wide or application-level histories for audit, accountability, and policy reasoning~\cite{bates2015trustworthy,pasquier2017practical}. These systems share with RAC the view that security decisions should be grounded in mediated execution state rather than untrusted program text or user-level explanations. RAC applies this view at the agent-controller authorization boundary, checking whether each continuation remains no more permissive than the authorization basis carried through verified lineage.

\subsection{Agent Benchmarks and Evaluation}

Tool-use and function-calling benchmarks evaluate whether LLM agents can select tools, call APIs, and complete tasks in both single-step and multi-step settings~\cite{qin2024toolllm,li2023apibank,patil2025berkeley,guo2024stabletoolbench}. Agent security benchmarks such as ToolEmu~\cite{ruan2024identifying}, AgentDojo~\cite{debenedetti2024agentdojo}, and InjecAgent~\cite{zhan2024injecagent} study unsafe tool behavior, prompt-injection robustness, and adversarial tool-integrated environments. Recent trace-level evaluation work shows that outcome-only scoring can miss procedural failures in agentic traces, and uses extracted behavioral rules to check trace compliance~\cite{sharma2026willful,paduraru2026trace}. In this paper, TraceBench serves a narrower role as an oracle-labeled specification workload for RAC's authorization-drift semantics. We complement it with high-confidence observable replay of planner-generated workflows and a live MCP filesystem case study that exercises pre-commit enforcement before server-side effects occur.

\section{Conclusion}

Agentic workflows increasingly execute user requests through runtime-expanded toolchains, where authorization risk can emerge from the composition of locally admissible steps. This paper formulated authorization drift as a workflow-level failure mode and presented RAC, a controller-side pre-commit guard that evaluates each pending continuation against the authorization basis inherited through accepted lineage. RAC leaves MCP servers unchanged and blocks unsupported continuations before they take effect. Evaluation with TraceBench, blind LLM-plan replay, practical baselines, a real MCP filesystem case study, and latency measurements shows that RAC implements the intended consistency semantics with low checking overhead in the prototype. RAC complements local authorization and agent-safety defenses by making inherited authorization support explicit at each workflow continuation.

% 可选：后面如需附录，再在模板里接 appendix 环境。

\appendix
\section{Proof Details for Theorem~1}
\label{app:proof-details}

This appendix expands the proof sketch in Section~\ref{sec:security-argument}. The proof is stated over the sequence of events that are allowed by RAC and committed through the monitored controller boundary.

\noindent\textbf{Lemma 1 (Event-field soundness).} If RAC constructs an authorization event $e_i$, then every security-critical field used by the checker is derived from an accepted source in Table~I.

\noindent\textit{Argument.} The event adapter constructs typed authorization events from controller-observed runtime metadata. Fields such as subject, action, resource identifier, purpose token, delegation state, runtime conditions, input anchors, and predecessor hints are accepted only when they are bound to the corresponding runtime source. Planner explanations, free-form tool outputs, tool self-reports, and agent-declared predecessor claims may be retained as diagnostic metadata, but they cannot create security-critical fields used by the checker.

\noindent\textbf{Lemma 2 (Origin soundness).} If $OriginOK$ holds for a pending event $e_i$, then every resource referenced by $e_i$ is supported by the session grant or by a verified predecessor anchor.

\noindent\textit{Argument.} Resource-origin verification accepts a resource identifier only when it belongs to the grant scope or appears in a verified input anchor. A verified anchor is hash-bound to a controller-observed output and carries only controller-verified resource identifiers. Therefore, a resource mentioned only in free-form text or unverified metadata cannot satisfy $OriginOK$.

\noindent\textbf{Lemma 3 (Lineage soundness).} If $LineageOK$ holds for a pending event $e_i$, then the predecessor basis loaded for $e_i$ belongs to an accepted same-session predecessor.

\noindent\textit{Argument.} RAC resolves predecessors from verified input anchors and lineage records. A predecessor is accepted only when the corresponding anchor producer is recorded in the same session, the anchor identifier and content hash match a verified output anchor in the lineage store, and the resolution is unambiguous under the prototype's single-predecessor discipline. Cross-session references, forged anchors, missing producers, and ambiguous predecessors therefore fail lineage resolution.

\noindent\textbf{Lemma 4 (Basis monotonicity).} For any admitted event $e_i$, the successor basis $B_i = Tighten(B_i^-, Obs(e_i))$ does not enlarge the authorization support carried by $B_i^-$.

\noindent\textit{Argument.} The tightening operator intersects the inherited basis with continuation profiles derived from the grant, manifest, and verified output anchors. When no narrower continuation profile is declared, the inherited basis is preserved. When a profile is declared, admissible actions, resources, purposes, and conditions are narrowed according to the profile. If a critical dimension becomes empty or incompatible, the event is blocked. Thus, tightening preserves or narrows the inherited authorization support.

\noindent\textbf{Lemma 5 (State exclusion for blocked events).} If RAC returns \textsc{Block} for a pending event $e_i$, then $e_i$ does not create successor lineage or successor basis state.

\noindent\textit{Argument.} The decision procedure writes lineage and basis state only on \textsc{Allow}. A blocked event is stopped before the external call is issued, and no accepted lineage record or finalized successor basis is appended for that event. Therefore, blocked events cannot become predecessor evidence for later continuations.

\noindent\textbf{Theorem 1 (Consistency of committed events).} Under the assumptions in Section~\ref{sec:threat-model}, every authorization-relevant event committed through RAC satisfies $Consistent(e_i)$ at the pre-commit boundary, and blocked events cannot become authorization evidence for later continuations.

\noindent\textit{Proof.} We prove the theorem by induction over the accepted sequence of committed authorization-relevant events. For the first accepted event, the inherited basis is initialized from the session grant. By Lemma~1, the event fields checked by RAC are controller-grounded. By the local access-control result, $LocalAllow(e_i)$ holds. By Lemma~2, $ValidOrigin(e_i)$ holds. The root case of Eq.~(\ref{eq:1}) holds for the initial grant-bound event; for continuations, Lemma~3 establishes $ValidLineage(e_i)$. Since RAC returns \textsc{Allow} only when the rule evaluator reports no violation, $NoMorePermissive(e_i,B_i^-)$ holds. Therefore, $Consistent(e_i)$ holds for the first committed event. For the induction step, assume all previously committed events satisfy Eq.~(\ref{eq:1}) and that their stored successor bases were produced by tightening. For the next committed event $e_i$, lineage resolution loads the basis from an accepted same-session predecessor by Lemma~3. The same reasoning establishes $LocalAllow(e_i)$, $ValidOrigin(e_i)$, $ValidLineage(e_i)$, and $NoMorePermissive(e_i,B_i^-)$. Lemma~4 ensures that the newly stored successor basis does not enlarge the inherited authorization support, and Lemma~5 ensures that blocked events cannot enter the accepted sequence. Hence, every committed event satisfies Eq.~(\ref{eq:1}), and rejected continuations cannot support later workflow steps.

\bibliographystyle{IEEEtran}
\bibliography{references}

\section*{LLM Usage Statement}

LLMs were used for editorial purposes in this manuscript, and all outputs were inspected by the authors to ensure accuracy and originality. LLMs were also used offline to generate candidate planner-style workflows for the observable replay study in Section~\ref{sec:rq2-llm-observable} and the deterministic filesystem replay corpus in Section~\ref{sec:rq3-mcp-filesystem}. These generated plans were normalized into fixed replay traces and replayed deterministically. The LLM was not part of RAC, the enforcement loop, or the reported decision procedure. Because live LLM generation can be nondeterministic and provider-dependent, the artifact evaluation uses the fixed normalized traces rather than requiring regeneration from a model endpoint. The high-confidence observable subset in RQ2 was derived from explicit tool-call facts, and all experimental configurations, reported results, and paper claims were reviewed by the authors.

\end{document}